\documentclass{article}
\usepackage{arxiv}
\usepackage[utf8]{inputenc} % allow utf-8 input
\usepackage[T1]{fontenc}    % use 8-bit T1 fonts
\usepackage{hyperref}       % hyperlinks
\usepackage{url}            % simple URL typesetting
\usepackage{booktabs}       % professional-quality tables
\usepackage{amsfonts}       % blackboard math symbols
\usepackage{nicefrac}       % compact symbols for 1/2, etc.
\usepackage{microtype}      % microtypography
\usepackage{lipsum}		
\usepackage{graphicx}
\usepackage{mathrsfs}
\usepackage{doi}
\usepackage{amsmath}
\usepackage[linesnumbered,ruled,vlined]{algorithm2e}
\usepackage{xcolor,soul}

\SetCommentSty{mycommfont}
\usepackage{caption,subcaption}
\allowdisplaybreaks
\usepackage{stmaryrd}
\usepackage{color,soul}
\usepackage{wrapfig}
\usepackage{cite}
\usepackage{cancel}
\hypersetup{
    colorlinks=true,
    linkcolor=blue,
    filecolor=magenta,      
    urlcolor=cyan,
}
\usepackage{courier}

\usepackage{listings}
\usepackage{xcolor}
\definecolor{codegreen}{rgb}{0,0.6,0}
\definecolor{codegray}{rgb}{0.5,0.5,0.5}
\definecolor{codepurple}{rgb}{0.58,0,0.82}
\definecolor{backcolour}{rgb}{0.95,0.95,0.92}

\lstdefinestyle{mystyle}{
  backgroundcolor=\color{backcolour},   commentstyle=\color{codegreen},
  keywordstyle=\color{magenta},
  numberstyle=\tiny\color{codegray},
  stringstyle=\color{codepurple},
  basicstyle=\ttfamily\footnotesize,
  breakatwhitespace=false,         
  breaklines=true,                 
  captionpos=b,                    
  keepspaces=true,                 
  numbers=left,                    
  numbersep=5pt,                  
  showspaces=false,                
  showstringspaces=false,
  showtabs=false,                  
  tabsize=2
}

\title{A machine learning accelerated FE$^2$ homogenization algorithm for elastic solids}

\author{
  {
  Saumik Dana}\\
	University of Southern California\\
	Los Angeles, CA 90007 \\
	\texttt{sdana@usc.edu} \\
 		\And
  {
  Mary F Wheeler} \\
	Oden Institute for Computational Engineering and Sciences\\
	University of Texas at Austin\\
	Austin, TX 78712 
}

\date{}

\begin{document}
\maketitle
\begin{abstract}
The FE$^2$ homogenization algorithm for multiscale modeling iterates between the macroscale and the microscale represented by a representative volume element, till convergence is achieved at every increment of macroscale loading. The information exchange between the two scales occurs at the gauss points of the macroscale finite element discretization. The microscale problem is also solved using finite elements on-the-fly thus rendering the algorithm computationally expensive for complex microstructures. We invoke machine learning to establish the input-output causality of the RVE boundary value problem using a neural network framework. This renders the RVE as a blackbox which gets the information from the macroscale as an input and gives information back to the macroscale as output, thereby eliminating the need for on-the-fly finite element solves at the RVE level. This framework has the potential to significantly accelerate the FE$^2$ algorithm.
\keywords{Machine learning \and FE$^2$ homogenization}
\end{abstract}
%%%
\section{Introduction}
\begin{figure}[htb!]
\centering
\includegraphics[scale=0.85]{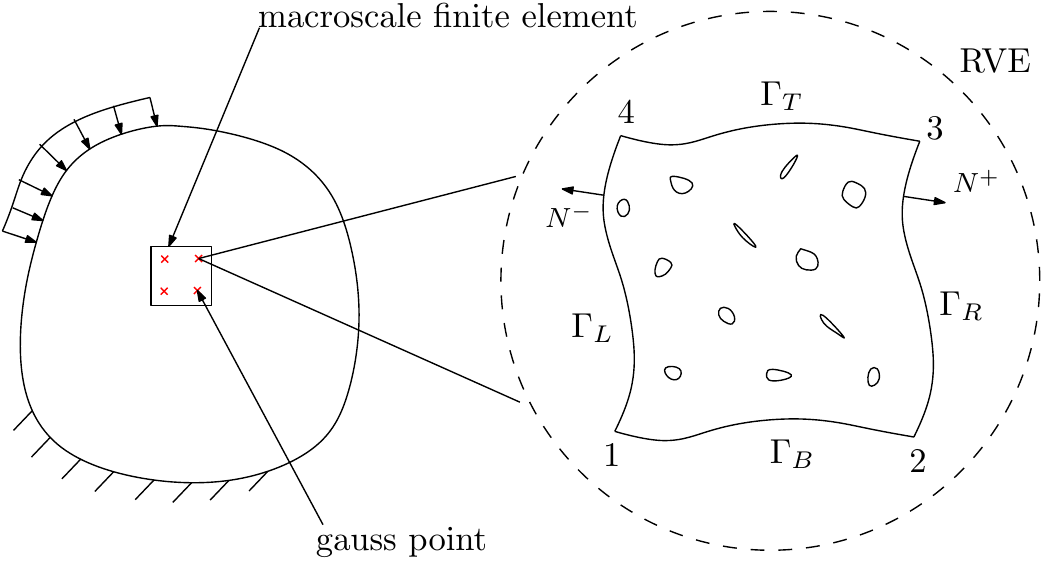}
\caption{A 2D depiction of the FE$^2$ algorithmic framework. The macroscale boundary value problem is discretized into finite elements. The gauss point level computations for the macroscale BVP work in conjunction with RVE scale solve corresponding to each gauss point.}
\label{dep}
\end{figure}
The RVE concept~\cite{hashin-1962,hill-1963,hill-1972,hashin-1983} is commonly used in the manufacturing sector to avoid using computationally expensive simulation platforms necessary to capture microstructural features. In essence, the features are captured in the RVE and averaged out over the RVE before any discretization technique is employed at the macroscale with the averaged properties as parameters. More often than not, a number of simulations are run with different microstructures and the statistical mean of the results from those simulations on the macroscale are used as guiding principles for the design of the part. The reason for running multiple simulations each with a different microstructure is that the microstructure is only known stochastically and not deterministically. The popular $FE^2$ numerical homogenization algorithm~\cite{geers,ozdemir-2008,schroder-2014} is commonly employed in which each gauss point for the finite element calculations at the macroscale is associated with a RVE and the information exchange between the two scales occurs at each of those gauss points via the deformation gradient. A 2D depiction of the algorithmic framework is given in Figure \ref{dep}. The reason for calling the framework $FE^2$ is that both the macroscale and the RVE scale are solved using finite element method. The information exchange between the two scales would need to occure multiple times at every increment of macro load to satisfy the accuracy and precision requirements expected of any numerical algorithm designed to solve a set of partial differential equations. The number of finite element solves at the RVE scale are proportional to the number of gauss points corresponding to the macroscale finite element mesh. This algorithmic framework is similar to staggered solution algorithms in which the coupled system of equations is decoupled at the PDE level using a constraint, and then the decoupled set of equations are solved sequentially and iteratively until convergence at a time step~\cite{dana-2018,dana2019design,dana2019system,dana2020,dana2020efficient,dana2021,danacg,danacmame,danathesis,jammoul2019RSC,JAMMOUL2020}. Depending on the complexity of the microstructure, the RVE solve itself would entail a lot of finite elements to resolve all the features in the RVE. The cumbersome computational cost would make the algorithm infeasible for complex microstructures. In lieu of that, methods to accelerate the algorithm need to be devised. One potential feature that can be incorporated in the algorithm is the use of neural network~\cite{rumelhart,nguyen1990neural} to establish the input-output causality of the RVE boundary value problem prior to any finite element solve at the macroscale. This would eliminate the need to solve the RVE boundary value problem on-the-fly as the neural network can be used as a blackbox which gets the information from the macroscale as an input and gives the information that the macroscale needs as an output. The elimination of the on-the-fly RVE solve would substantially reduce the computational burden on the algorithmic framework. In essence, the FE$^2$ framework would effectively be converted to a $FE^1$ framework since it would require a finite element solve only at the macroscale. It is now important to identify what information is provided to the RVE from the macroscale and what information is provided back to the macroscale by the RVE. In case of elastic solids, the information exchange is as follows
\begin{align*}
\mathrm{Macroscale} \xrightarrow{\mathrm{deformation\,\,gradient}} \mathrm{RVE} \xrightarrow{\mathrm{homogenized\,\,stress\,\,measure}} \mathrm{Macroscale}
\end{align*} 
\par
The concepts of deformation gradient and the particular stress measure are explained in Appendix \ref{explain}. The deformation gradient manifests itself as boundary conditions on the RVE. Periodic boundary conditions satisfy the Hill-Mandel condition~\cite{hillmandel1,hazanov-1998} of energetic equivalence between the two scales and are generally the optimal choice from the standpoint of macroscale accuracy~\cite{michel-1999,kouznetsova2001,miehe-2003,dijk-2015}. The imposition of periodic boundary conditions are explained in Appendix \ref{perbcmod}.
%%%
\section{The machine learning aspect}
\begin{figure}[htb!]
\centering
\includegraphics[scale=0.85]{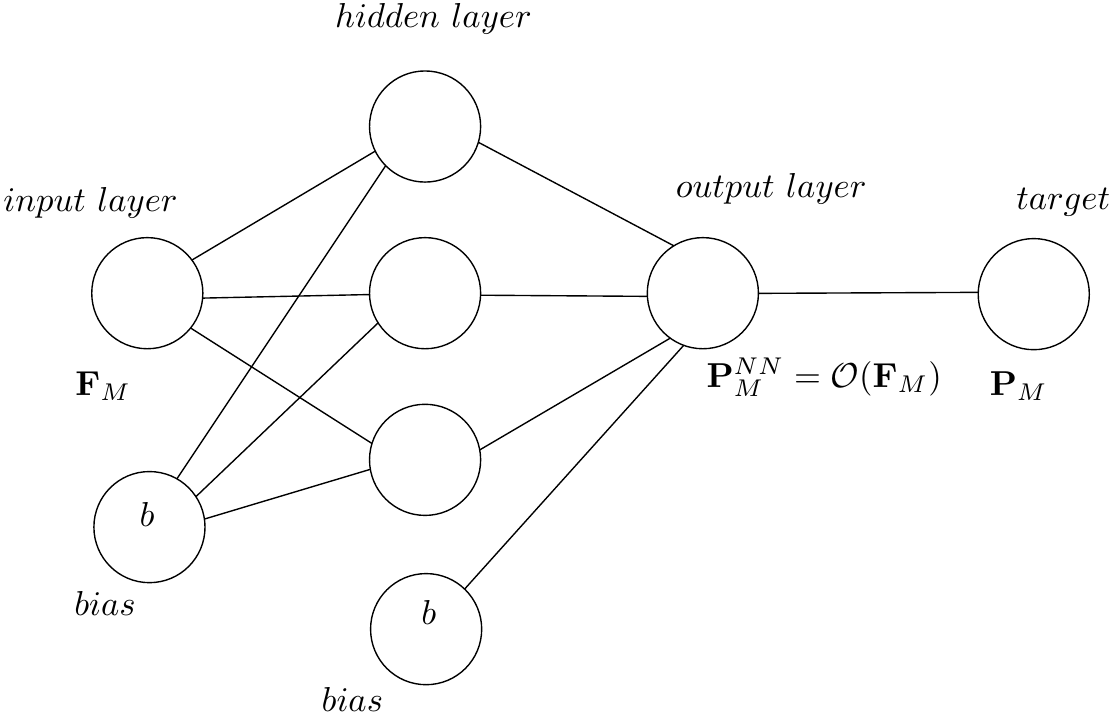}
\caption{A 1-3-1 neural network with macroscale deformation gradient as input and homogenized first P-K stress as the target output.}
\label{depnn}
\end{figure}
We follow the outline laid out in \cite{mlpeter} to incorporate machine learning in the algorithmic framework. A neural network is composed of several connected layers of artificial neurons and biases where the data is fed into the input layer and flows through some hidden layers. The output is predicted in the output layer. The neurons from different layers are connected through weights $w$. In the data collection phase, the data flows in one way from the input layer to the target. A simple 1-3-1 neural network with deformation gradient as input and homogenized first P-K stress as target is depicted in Figure \ref{depnn}.
At each neuron, an activation function is attached. The output of each neuron is computed by multiplying the outputs from the previous layer with the corresponding weights. For the neuron $j$ in layer $k$, the data from the previous layer $k-1$ is summed up and then altered by an activation function. The output of neuron $j$ in layer $k$ is computed as
\begin{align*}
o_j^k=\mathcal{F}(\sum\limits_{i=1}^N w_{ij}o_i^{k-1}+b_i^{k-1})
\end{align*}
where $N$ is the number of neurons in the previous layer $k-1$, $w_{ij}$ is the weight connecting neurons $i$ and $j$, $o_i^{k-1}$ is the output of neuron $i$ in layer $k-1$ and $b_i^{k-1}$ is its bias. A common choice for the activation function is the sigmoid
\begin{align*}
\mathcal{F}=\frac{2}{1+e^{-2x}}-1
\end{align*}
In the training phase, the weights of neural network will be initialized firstly, (see \cite{nguyen}), which is followed by the weights updating using a training algorithm such that the global error is minimized. The global error, also named as loss function or network performance, is defined according to the difference between the network prediction and the target data. To minimize the global error, the Levenberg-Marquardt algorithm (see \cite{hagan}) is applied to update the weights. The steps in the machine learning addendum to the algorithm are
\begin{itemize}
\item The RVE boundary value problem is solved using finite elements for a myriad of deformation gradients with the non-linear neo-Hookean model as the stress-strain relation
\begin{align*}
\boldsymbol{\sigma}=\frac{1}{2}\frac{\lambda}{J}(J^2-1)\mathbf{I}+\frac{\mu}{J}(\mathbf{b}-\mathbf{I})
\end{align*}
The deformation gradient is fed to the RVE problem via periodic boundary conditions as explained in Appendix \ref{perbcmod}. The homogenized first P-K stress is obtained for each of these deformation gradients using the relationship \eqref{four}.  
\item This data is then used to build the input-output causality as follows
\begin{align}
\label{map}
\mathbf{P}_M^{NN}=\mathcal{O}(\mathbf{F}_M)
\end{align}
where $\mathcal{O}$ is the map between the deformation gradient and the output of the neural network.
\end{itemize}
%%%
\section{Algorithmic framework in a nutshell}
In the initialization stage,
\begin{itemize}
\item Establish the input-output causality of the RVE boundary value problem as in \eqref{map}
\end{itemize}
Once the initalization phase is complete, 
\begin{itemize}
\item An increment of macro load is applied
\item Macroscale BVP is solved using the macroscale stiffness computed in \eqref{seven} 
\item The macroscale deformation gradient is updated
\item Periodic boundary conditions are imposed on RVE in accordance with \eqref{perbc}
\item Homogenized first P-K stress is obtained in accordance with \eqref{map}
\item The gauss point level homogenized first P-K stress is used to compute internal forces at macroscale finite element nodes
\end{itemize}
If these internal forces are in balance with the prescribed macro load, incremental convergence has been achieved and steps $1-6$ are repeated. 
If that is not the case, steps $2-6$ are repeated. 
%%%
\begin{algorithm}[htb!]
%\begin{algorithmic}
\caption{Machine learning based FE$^2$ homogenization for elastic solids}
\textit{Use machine learning to establish the relationship \eqref{map}}\;
$\mathbf{F}_M\gets \mathbf{I}$\tcp*{Initialize deformation gradient}
\For{$E \in \mathscr{T}_h$}
{
\tcc{Loop over macroscale finite elements}
\For{$g \in \mathscr{G}$}
{
\tcc{Loop over gauss points}
 RVE $\leftrightarrow\,\,g$\tcp*{Assign a RVE to each gauss point}
 Discretize the RVE
 Calculate homogenized macroscopic tangent stiffness in accordance with \eqref{seven}
}
 Assemble macroscopic tangent stiffness over gauss points\;
}
 Assemble macroscopic tangent stiffness over finite elements\;
\While{$t\le T$}
 {
 Apply increment of macro load\;
\While{(Internal force-Macro load $>$ TOL)} 
{
\tcc{At macroscale finite element nodes}
 Solve macroscale problem for $\delta \mathbf{F}_M$\;
 $\mathbf{F}_M\gets \mathbf{F}_M+\delta \mathbf{F}_M$ \tcp*{Update deformation gradient}
\For{$E \in \mathscr{T}_h$} 
{
\tcc{Loop over macroscale finite elements}
\For{$g \in \mathscr{G}$}
{
\tcc{Loop over gauss points}
 Prescribe periodic BCs in accordance with \eqref{perbc}\; 
 \st{\textit{Solve RVE problem}}\tcp*{\textit{Not needed as the relationship \eqref{map} has been estalished using machine learning}}
 Calculate first P-K stress in accordance with \eqref{map}\;
}
}
Compute internal forces at finite element nodes\;
}
}
%\end{algorithmic}
\end{algorithm}
%%%
\section*{Conflict of interest}
The authors declare that they have no conflict of interest.
%%%%
\appendix
\section{The deformation gradient and first P-K stress}\label{explain}
\begin{figure}[htb!]
\centering
\includegraphics[scale=0.9]{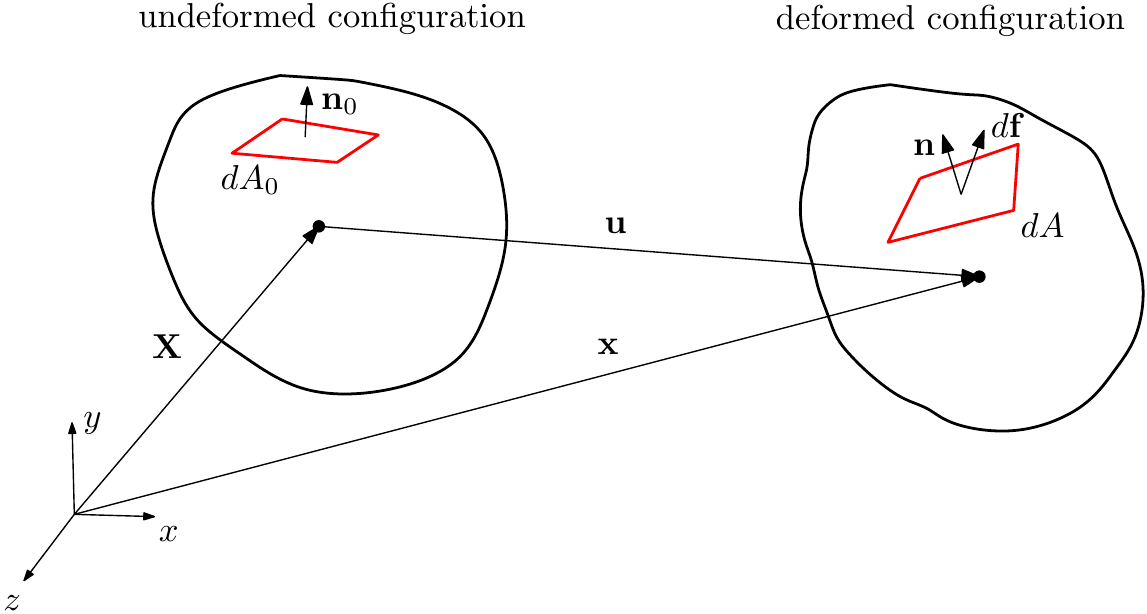}
\caption{$\mathbf{X}$ is position vector of point in reference configuration and $\mathbf{x}=\mathbf{X}+\mathbf{u}$ is the position vector the same point in the deformed configuration. Meanwhile, an elemental area $dA_0$ with unit normal $\mathbf{n}_0$ deforms to $dA$ with unit normal $\mathbf{n}$ under the transformation.}
\label{defgrad}
\end{figure}
As shown in Figure \ref{defgrad}, let $\mathbf{u}$ be the macroscale deformation field. 
The macroscale deformation gradient $\mathbf{F}_M$ is the macroscale spatial derivative of $\mathbf{x}$ in the reference configuration as follows
\begin{align*}
\mathbf{F}=\mathbf{x} \otimes \nabla_X \equiv \mathbf{I}+\mathbf{u}\otimes \nabla_X
\end{align*}
 An incremental force $d\mathbf{f}$ is defined with respect to the Cauchy stress $\boldsymbol{\sigma}$ and the first Piola-Kirchoff stress $\mathbf{P}$ in the deformed and reference configurations respectively as follows
\begin{align*}
d\mathbf{f}=\boldsymbol{\sigma}\mathbf{n}\,dA=\mathbf{P}\mathbf{n}_0\,dA_0
\end{align*}
%%%%
\section{Periodic boundary conditions on RVE}\label{perbcmod}
\begin{figure}[htb!]
\centering
\includegraphics[scale=1.0]{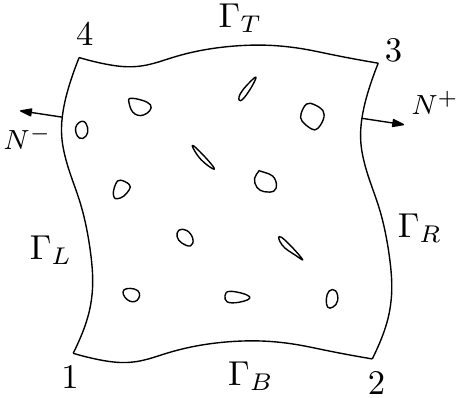}
\caption{Typical 2D RVE with pertinent microstructural features. $\Gamma_L$ and $\Gamma_R$ are mirror images so are $\Gamma_T$ and $\Gamma_B$. This helps in easy implementation of periodic boundary conditions on the RVE in accordance with \cite{kouznetsova2001}.}
\label{per}
\end{figure}
The typical RVE for imposition of periodic boundary conditions is shown in Figure \ref{per}. After each macroscale BVP solve, the deformation gradient is updated and the new position vectors of the vertices of the RVE are obtained using
\begin{align}
\label{perbc}
\mathbf{x}=\mathbf{F}_M \mathbf{X}
\end{align}
where $\mathbf{X}$ represents position vector in the reference configuration. This alongwith the shape periodicity of the RVE enables the implementation of periodic boundary conditions on RVE. It is easy to see that the prescribed periodic boundary conditions are Dirichlet boundary conditions. 
%%%
\section{Computation of homogenized first P-K stress at the macroscale}\label{macro}
The linear momentum balance for the macroscale BVP in the reference configuration is given by
\begin{align*}
\nabla_X \cdot \mathbf{P}_M+\mathbf{b}=\mathbf{0}
\end{align*}
where $\mathbf{b}$ is the body force vector. The macroscale incremental constitutive law is
\begin{align}
\label{zero}
\delta \mathbf{P}_M=\mathbb{C}_M\delta \mathbf{F}_M
\end{align}
where $\mathbb{C}_M$ is the fourth order macroscale material property tensor.
The determination of $\mathbb{C}_M$ proceeds as follows: First, the RVE scale linear momemtum balance is expressed in the indicial notation as
\begin{align*}
P_{ik,k}+b_i=0\qquad i,k=1,2,3
\end{align*}
where the notation $(\cdot)_{,k}\equiv \frac{\partial (\cdot)}{\partial X_k}$ is used to denote the spatial derivative in the reference configuration. 
%as follows 
%\begin{align*}
%(\cdot)_{,k}\equiv \frac{\partial (\cdot)}{\partial X_k}
%\end{align*}
Before we proceed, we assume that the body force is zero, and obtain the following using chain rule for differentiation 
\begin{align}
\label{one}
(P_{ik}X_j)_{,k}=P_{ik,k}X_j+P_{ik}\delta_{jk}=-\cancelto{0}{b_i} X_j+P_{ij}
\end{align}
We express the macroscale first P-K stress in indicial notation as follows
\begin{align}
\label{two}
P_{M_{ij}}=\frac{1}{V_0}\int\limits_{V_0}P_{ij}\,dV_0=\frac{1}{V_0}\int\limits_{V_0}(P_{ik}X_j)_{,k}\,dV_0=\frac{1}{V_0}\int\limits_{\Gamma_0}P_{ik}n_{0_k} X_j \,d\Gamma_0
\end{align}
where the third equality follows from \eqref{one} and the fourth equality follows from divergence theorem.
We then write \eqref{two} in tensorial notation as
\begin{align}
\label{three}
\mathbf{P}_M=\frac{1}{V_0}\int\limits_{\Gamma_0}\mathbf{t}_0\otimes \mathbf{X}\,d\Gamma_0
\end{align}
We know that the RVE level BVP is also solved using finite elements. Let $N_p$ be the number of boundary nodes for the RVE scale discretized domain and let $\mathbf{f}_p^{(i)}$ be the force on $i^{th}$ boundary node. We can rewrite \eqref{three} as 
\begin{align}
\label{four}
\mathbf{P}_M=\frac{1}{V_0}\int\limits_{\Gamma_0}\mathbf{t}_0\otimes \mathbf{X}\,d\Gamma_0=\frac{1}{V_0}\sum\limits_{i=1}^{N_p}\mathbf{f}_p^{(i)} \otimes \mathbf{X}^{(i)}
\end{align}
%%%%
\section{Computation of homogenized tangent stiffness at macroscale}
Let $\mathbf{u}_f$ represent the displacement DOFs corresponding to the interior nodes and $\mathbf{u}_p$ represent the displacement DOFs corresponding to the boundary nodes. The force displacement relation for the RVE scale problem is
\begin{align*}
\cancelto{\mathbf{K}^{RVE}}{\begin{bmatrix}
\mathbf{K}_{pp} & \mathbf{K}_{pf}\\
\mathbf{K}_{fp} & \mathbf{K}_{ff}
\end{bmatrix}}\left\{\begin{array}{c}
\delta \mathbf{u}_p\\
\delta \mathbf{u}_f
\end{array} \right\}=\left\{\begin{array}{c}
\delta \mathbf{f}_p\\
\mathbf{0}
\end{array} \right\}
\end{align*}
where the matrix $\mathbf{K}^{RVE}$ is dictated by the microstructure and is known apriori. We knock off DOFs corresponding to internal nodes to obtain
\begin{align}
\label{five}
[\cancelto{\mathbf{K}}{\mathbf{K}_{pp}-\mathbf{K}_{pf}(\mathbf{K}_{ff})^{-1}\mathbf{K}_{fp}}]\{\delta \mathbf{u}_p\}=\{\delta \mathbf{f}_p\}
\end{align}
The incremental macroscopic first PK stress is obtained as
\begin{align}
\nonumber
\delta \mathbf{P}_M&=\frac{1}{V_0}\sum\limits_{i=1}^{N_p}\delta \mathbf{f}_p^{(i)} \otimes \mathbf{X}^{(i)}\qquad (from\,\eqref{four})\\
\nonumber
&=\frac{1}{V_0}\sum\limits_{i=1}^{N_p}\sum\limits_{j=1}^{N_p}\mathbf{K}^{(ij)}\delta \mathbf{u}_p^{(j)} \otimes \mathbf{X}^{(i)}\qquad (from\,\eqref{five})\\
\label{six}
&=\frac{1}{V_0}\sum\limits_{i=1}^{N_p}\sum\limits_{j=1}^{N_p}\mathbf{K}^{(ij)}\delta \mathbf{F}_M \mathbf{X}^{(j)}\otimes \mathbf{X}^{(i)}\qquad (\delta \mathbf{u}=\delta \mathbf{F}_M \mathbf{X})
\end{align}
Comparing \eqref{six} with \eqref{zero}, we get
\begin{align}
\label{seven}
\mathbb{C}_{M_{abcd}}= \frac{1}{V_0}\sum\limits_{i=1}^{N_p}\sum\limits_{j=1}^{N_p}\mathbf{K}_{ac}^{(ij)}\mathbf{X}^{(i)}_b\mathbf{X}^{(i)}_d\qquad a,b,c,d=1,2,3
\end{align}
%%%%
\bibliographystyle{unsrt}      
\bibliography{diss}

\begin{thebibliography}{10}

\bibitem{hashin-1962}
Z.~Hashin and S.~Shtrikman.
\newblock On some variational principles in anisotropic and nonhomogeneous
  elasticity.
\newblock {\em Journal of the Mechanics and Physics of Solids}, 10(4):335--342,
  1962.

\bibitem{hill-1963}
R.~Hill.
\newblock Elastic properties of reinforced solids: Some theoretical principles.
\newblock {\em Journal of the Mechanics and Physics of Solids}, 11(5):357--372,
  1963.

\bibitem{hill-1972}
R.~Hill.
\newblock On constitutive macro-variables for heterogeneous solids at finite
  strain.
\newblock {\em Proceedings Mathematical Physical and Engineering Sciences},
  326(1565):131--147, 1972.

\bibitem{hashin-1983}
Z.~Hashin.
\newblock Analysis of composite materials--a survey.
\newblock {\em Journal of Applied Mechanics}, 50(3):481--505, 1983.

\bibitem{geers}
M.G.D. Geers, V.G. Kouznetsova, and W.A.M. Brekelmans.
\newblock Multi-scale computational homogenization: Trends and challenges.
\newblock {\em Journal of Computational and Applied Mathematics}, 234(7):2175
  -- 2182, 2010.
\newblock Fourth International Conference on Advanced Computational Methods in
  Engineering (ACOMEN 2008).

\bibitem{ozdemir-2008}
I.~Ozdemir, W.~A.~M. Brekelmans, and M.~G.~D. Geers.
\newblock Fe2 computational homogenization for the thermo-mechanical analysis
  of heterogeneous solids.
\newblock {\em Computer Methods in Applied Mechanics and Engineering},
  198(3):602 -- 613, 2008.

\bibitem{schroder-2014}
J{\"o}rg Schr{\"o}der.
\newblock A numerical two-scale homogenization scheme: the fe2-method.
\newblock In {\em Plasticity and beyond}, pages 1--64. Springer, 2014.

\bibitem{dana-2018}
Saumik Dana, Benjamin Ganis, and Mary~F. Wheeler.
\newblock A multiscale fixed stress split iterative scheme for coupled flow and
  poromechanics in deep subsurface reservoirs.
\newblock {\em Journal of Computational Physics}, 352:1--22, 2018.

\bibitem{dana2019design}
Saumik Dana and Mary~F Wheeler.
\newblock Design of convergence criterion for fixed stress split iterative
  scheme for small strain anisotropic poroelastoplasticity coupled with single
  phase flow.
\newblock {\em arXiv preprint arXiv:1912.06476}, 2019.

\bibitem{dana2019system}
Saumik Dana.
\newblock System of equations and staggered solution algorithm for immiscible
  two-phase flow coupled with linear poromechanics.
\newblock {\em arXiv preprint arXiv:1912.04703}, 2019.

\bibitem{dana2020}
Saumik Dana, Joel Ita, and Mary~F Wheeler.
\newblock The correspondence between voigt and reuss bounds and the decoupling
  constraint in a two-grid staggered algorithm for consolidation in
  heterogeneous porous media.
\newblock {\em Multiscale Modeling \& Simulation}, 18(1):221--239, 2020.

\bibitem{dana2020efficient}
Saumik Dana and Mary~F Wheeler.
\newblock An efficient algorithm for numerical homogenization of fluid filled
  porous solids: part-i.
\newblock {\em arXiv preprint arXiv:2002.03770}, 2020.

\bibitem{dana2021}
Saumik Dana, Xiaoxi Zhao, and Birendra Jha.
\newblock Two-grid method on unstructured tetrahedra: Applying computational
  geometry to staggered solution of coupled flow and mechanics problems.
\newblock {\em arXiv preprint arXiv:2102.04455}, 2021.

\bibitem{danacg}
S.~Dana and M.~F. Wheeler.
\newblock Convergence analysis of fixed stress split iterative scheme for
  anisotropic poroelasticity with tensor biot parameter.
\newblock {\em Computational Geosciences}, 22(5):1219--1230, 2018.

\bibitem{danacmame}
S.~Dana and M.~F. Wheeler.
\newblock Convergence analysis of two-grid fixed stress split iterative scheme
  for coupled flow and deformation in heterogeneous poroelastic media.
\newblock {\em Computer Methods in Applied Mechanics and Engineering},
  341:788--806, 2018.

\bibitem{danathesis}
S.~Dana.
\newblock {\em Addressing challenges in modeling of coupled flow and
  poromechanics in deep subsurface reservoirs}.
\newblock PhD thesis, The University of Texas at Austin, 2018.

\bibitem{jammoul2019RSC}
Mohamad Jammoul, Benjamin Ganis, and Mary~F. Wheeler.
\newblock General semi-structured discretization for flow and geomechanics on
  diffusive fracture networks.
\newblock In {\em SPE Reservoir Simulation Conference}. Society of Petroleum
  Engineers, 2019.

\bibitem{JAMMOUL2020}
Mohamad Jammoul, Mary~F. Wheeler, and Thomas Wick.
\newblock A phase-field multirate scheme with stabilized iterative coupling for
  pressure driven fracture propagation in porous media.
\newblock {\em Computers {\&} Mathematics with Applications}, 2021.

\bibitem{rumelhart}
David~E Rumelhart, Geoffrey~E Hinton, and Ronald~J Williams.
\newblock Learning internal representations by error propagation.
\newblock Technical report, California Univ San Diego La Jolla Inst for
  Cognitive Science, 1985.

\bibitem{nguyen1990neural}
Derrick~H Nguyen and Bernard Widrow.
\newblock Neural networks for self-learning control systems.
\newblock {\em IEEE Control systems magazine}, 10(3):18--23, 1990.

\bibitem{hillmandel1}
R.~Hill.
\newblock A self-consistent mechanics of composite materials.
\newblock {\em Journal of the Mechanics and Physics of Solids}, 13(4):213--222,
  1965.

\bibitem{hazanov-1998}
S.~Hazanov.
\newblock Hill condition and overall properties of composites.
\newblock {\em Archive of Applied Mechanics}, 68(6):385--394, 1998.

\bibitem{michel-1999}
J.~C. Michel, H.~Moulinec, and P.~Suquet.
\newblock Effective properties of composite materials with periodic
  microstructure: a computational approach.
\newblock {\em Computer Methods in Applied Mechanics and Engineering},
  172:109--143, 1999.

\bibitem{kouznetsova2001}
V.~V.~Kouznetsova, W.~A.~M. Brekelmans, and F.~P.~T. Baaijens.
\newblock An approach to micro-macro modeling of heterogeneous materials.
\newblock {\em Computational Mechanics}, 27(1):37--48, 2001.

\bibitem{miehe-2003}
C.~Miehe.
\newblock Computational micro-to-macro transitions for discretized
  micro-structures of heterogeneous materials at finite strains based on the
  minimization of averaged incremental energy.
\newblock {\em Computer Methods in Applied Mechanics and Engineering},
  192:559--591, 2003.

\bibitem{dijk-2015}
N.~P. van Dijk.
\newblock Formulation and implementation of stress- and/or strain-driven
  computational homogenization for finite strain.
\newblock {\em International Journal for Numerical Methods in Engineering},
  pages 1009--1028, 2015.

\bibitem{mlpeter}
Dengpeng Huang, Jan~Niklas Fuhg, Christian Wei{\ss}enfels, and Peter Wriggers.
\newblock A machine learning based plasticity model using proper orthogonal
  decomposition.
\newblock {\em arXiv preprint arXiv:2001.03438}, 2020.

\bibitem{nguyen}
Derrick Nguyen and Bernard Widrow.
\newblock Improving the learning speed of 2-layer neural networks by choosing
  initial values of the adaptive weights.
\newblock In {\em 1990 IJCNN International Joint Conference on Neural
  Networks}, pages 21--26. IEEE, 1990.

\bibitem{hagan}
Martin~T Hagan and Mohammad~B Menhaj.
\newblock Training feedforward networks with the marquardt algorithm.
\newblock {\em IEEE transactions on Neural Networks}, 5(6):989--993, 1994.

\end{thebibliography}

\end{document}